\input{psfig}
\documentstyle[preprint,prl,aps]{revtex}

\begin{document}


\title{Amplitude equations and pattern selection in Faraday waves}
\author{Peilong Chen$^{1}$ and Jorge Vi\~nals$^{1,2}$} 
\address{$^{1}$ Supercomputer Computations Research Institute,
	 Florida State University,
    	 Tallahassee, Florida 32306-4052, 
	$^{2}$ Department of Chemical Engineering,
	FAMU-FSU College of Engineering,
	Tallahassee, Florida 32310}
\date{\today}
\maketitle
\begin{abstract}
We present a systematic nonlinear theory of pattern selection for
parametric surface waves (Faraday waves), not restricted to fluids of
low viscosity. A standing wave amplitude equation is derived from the
Navier-Stokes equations that is of gradient form.  The associated
Lyapunov function is calculated for different regular patterns to
determine the selected pattern near threshold. For fluids of large
viscosity, the selected wave pattern consists of parallel stripes. At
lower viscosity, patterns of square symmetry are obtained in the
capillary regime (large frequencies).  At lower frequencies (the mixed
gravity-capillary regime), a sequence of six-fold (hexagonal), eight-fold,
$\ldots$ patterns are predicted. The regions of stability of the
various patterns are in quantitative agreement with recent experiments
conducted in large aspect ratio systems.

\end{abstract}


Parametrically driven surface waves (also known as Faraday waves)
appear on the free surface of fluid layer which is periodically
vibrated in the direction normal to the surface at rest.  Above a
certain critical value of the driving amplitude, the planar surface
becomes unstable to a pattern of standing waves \cite{review}. If the
viscosity of the fluid is large, the bifurcating wave pattern consists
of parallel stripes. At lower viscosity, patterns of square symmetry
are observed in the capillary regime (large frequencies)
\cite{square}. At lower frequencies (the mixed gravity-capillary
regime), hexagonal, eight-fold, and ten-fold patterns have been
observed \cite{patterns,kudrolli96a,binks96}. We present a weakly
nonlinear analysis of the equations governing fluid motion that
predicts stationary wave patterns with these
symmetries. Their boundaries of stability agree quantitatively with
experiments. \cite{kudrolli96a,binks96}

Pattern selection in confined geometries can be often understood in
terms of the spatial modes of the base state that become linearly
unstable and the geometry of the system. Extended systems, on the
other hand, allow a richer nonlinear competition of linearly unstable
modes, in part due to the restoration of some of the symmetries of the
original system that had been broken by the boundaries. The
distinction can be further quantified by introducing the coherence
length of the pattern $\xi$, the linear size of the system $L$, and
the characteristic wavelength of the pattern $1/k_{0}$. Directional
solidification from the melt is a typical example of a system in the
limit $\xi_{0} \ll 1/k_{0}\ll L$, fact that follows from the extremely
flat neutral stability curve \cite{re:mullins64}. Experiments often
show a narrow distribution of wavelengths in the stationary state, the
prediction of which has proven elusive. Attempts at deriving amplitude
equations valid near onset have not been successful precisely because
of the condition $\xi_{0} \ll 1/k_{0}$. A typical and widely studied
example of the intermediate range concerns Rayleigh-B\'enard
convection in large aspect ratio cells $1/k_{0} \precsim \xi \ll
L$. Although amplitude equations predict the existence of stable
parallel rolls above onset, such a state is not generically observed
in fluids of low or moderate Prandtl number except under carefully
prepared initial conditions. Instead, a spatio-temporally chaotic
state emerges that has been termed spiral defect chaos \cite{spiral}.
The role that long wavelength modes (or mean flow) play in the
development of such a state is still a matter of research. Faraday
waves, on the other hand, are a prototypical case of a system with a
large coherence length $\xi \gg 1/k_{0}$, and therefore one would
expect that amplitude equations would allow quantitative predictions
of pattern selection near onset. Furthermore, the
physical system and the experimental conditions are completely
determined by a few independent parameters that can be obtained with
reasonable accuracy.

Progress in deriving suitable amplitude equations for Faraday waves has
proved difficult. Since Faraday waves are almost a Hamiltonian system
(weakly dissipative), most analyses are based on a hamiltonian
description for the ideal (inviscid) limit, and treat
viscous or dissipative effects as a perturbation
\cite{milner91,miles93}.  The central question that arises in this
case concerns the origin of any nonlinear saturation of the wave
pattern. First, general symmetry considerations for Hamiltonian
systems (\lq\lq chiral symmetry") prohibit cubic terms in standing
wave amplitude equations \cite{review,re:coullet94}. In addition,
it has been argued that linear viscous terms in the original equations
governing Faraday waves only contribute linear damping terms to the
amplitude equations \cite{milner91}. These observations have 
contributed to the belief that saturation in near-Hamiltonian systems 
generically occurs
through either nonlinear (cubic) viscous terms or fifth order
hamiltonian terms, both of which are very difficult to obtain
explicitly \cite{milner91}. The issue, however, has remained somewhat
controversial for the case of Faraday waves \cite{miles93}, and we
mention, for example, two recent studies that considered
non-dissipative cubic terms in the amplitude equation. M\"{u}ller
\cite{muller94} has shown that the generic form of a cubic amplitude 
equation would allow patterns of standing waves with the observed symmetries, 
given suitable choices of the coefficients. A more recent analysis
\cite{zhang96} explicitly addressed these issues in the limit of small
fluid viscosity. A set of quasi-potential equations was derived by
considering the rotational flow within a small viscous layer near the
free surface, and assuming potential flow in the bulk. A standing wave
amplitude equation was obtained with cubic order terms given
explicitly.

Although this latter model predicted bifurcations to square and
higher-symmetry patterns, it did rely on an uncontrolled approximation
concerning nonlinear viscous terms. As a consequence, its region of
validity is difficult to asses. In particular, it failed to yield the
observed stripe patterns at intermediate and large viscosities.  We
present here a general calculation that overcomes these difficulties,
and that leads to the experimentally observed regular (periodic or
quasi-periodic) standing wave patterns above onset. As part of the
derivation, we also obtain an analytical expression for the linear
threshold of instability, which was previously known only numerically
\cite{kumar94}. Not being confined to small viscosities is also
important for comparison with experiments since in this case it is
easier to achieve the large aspect ratio limit, and hence to study
pattern formation without the influence of side walls.


We consider an incompressible viscous fluid under vertical vibration
$f\cos(\omega t)$ of amplitude $f$, and angular frequency $\omega$.
The fluid at rest has a free surface at $z=0$, extends to $z=-\infty$,
and is unbounded in the $x$--$y$ direction. The equation governing
fluid motion is
$$
  \partial_t{\bf u} + ({\bf u}\cdot{\bf\nabla}){\bf u}
 = -\frac{1}{\rho}{\bf\nabla}p + \nu\nabla^2{\bf u} + G(t){\hat{\bf e}}_z,
$$
with ${\bf u}$ the velocity field, $p$ the pressure, $\rho$ and $\nu$
the density and kinematic viscosity of the fluid respectively, and
$G(t) = - g - {1 \over 2}f(e^{i\omega t} +e^{-i\omega t})$ the
effective gravity. For $f$ below the threshold of instability, the
base state is ${\bf u}=0$ and $p=\rho G(t)z$. We first eliminate the
explicit dependence on the pressure by taking
$-{\bf\nabla}\times{\bf\nabla}\times$ to obtain
$$
  \partial_t\nabla^2{\bf u} - \nu\nabla^2\nabla^2{\bf u}
 ={\bf\nabla}\times{\bf\nabla}\times({\bf u}\cdot{\bf\nabla}){\bf u}.
$$
Here the continuity equation ${\bf\nabla}\cdot{\bf u}=0$ has also 
been used.

The position of the free surface is denoted by $\zeta(x,y)$, the unit
normal ${\bf\hat n} = (-\partial_x\zeta,-\partial_y\zeta,1)$, and the two 
tangential unit vectors are ${\bf\hat t}_1 = (1,0,\partial_x\zeta)$ and 
${\bf\hat t}_2 = (0,1,\partial_y\zeta)$.
Besides the null conditions at $z=-\infty$, there are three boundary
conditions to be satisfied at the free surface,
\begin{eqnarray*}
  \partial_t\zeta + ({\bf u}\cdot{\bf\nabla}_H)\zeta &=& w|_{z=\zeta} 
    \nonumber \\
  {\bf\hat t}_m\cdot{\bf T}\cdot{\bf\hat n}|_{z=\zeta} &=& 0, \qquad m=1,2
    \nonumber \\
  {\bf\hat n}\cdot{\bf T}\cdot{\bf\hat n}|_{z=\zeta} &=& 2H\sigma,
\end{eqnarray*}
with ${\bf\nabla}_H\equiv {\bf\hat e}_x\partial_x + {\bf\hat
e}_y\partial_y$, ${\bf T}$ the stress tensor with components 
$ T_{ij} = [-p-\rho
G(t)z]\delta_{ij}+\rho\nu(\partial_ju_i+\partial_iu_j) $, $\sigma$ the
surface tension, and $2H$ the mean curvature of the free surface
\cite{lamb}.
  

First we consider the linear stability of a subharmonic standing
wave,
$$
  w_0 = \cos(kx) \sum_{j=1,3,5,\cdots} e^{ji\omega t/2} w_{0}^j(z)A_j 
                          +\hbox{c.c.},
$$
where $w_{0}$ is the $z$-component of the velocity field, and a
similar expansion for $\zeta_0$. Substitution into the linearized
equation of motion, $ \left( \partial_t\nabla^2 - \nu\nabla^2\nabla^2
\right) w_0 = 0$, and the linearized kinematic and tangential stress
boundary conditions, $\partial_t\zeta_0 - w_0=0$ and
$(\nabla_H^2-\partial_z^2)w_0=0$, we find
$$
  w_0^j(z)=\nu(k^2+q_j^2)e^{kz} -2\nu k^2e^{q_jz},
$$
with $q_j^2\equiv k^2+ji\omega/2\nu$. Note that the boundary
conditions at $z=\zeta$ have been expanded around $z=0$.
(Generalization to a finite fluid depth is straightforward: the term
$d_1e^{-kz} + d_2e^{-q_jz}$ will also be in $w_0^j(z)$ with $d_1$ and
$d_2$ to be determined by the additional boundary conditions at the
bottom.)

The critical amplitude $f_0$ is determined
by the linearized normal stress boundary condition, which is, after
using the momentum equation to eliminate $p_0$,
\begin{eqnarray*}
  2\rho\nu\nabla_H^2&\partial_z&w_0 - \rho\partial_t\partial_zw_0
 +\rho\nu\nabla^2\partial_zw_0
 +\rho g\nabla_H^2\zeta_0 \\
 &+&{\textstyle{1\over 2}}
  \rho f\left(e^{i\omega t}+e^{-i\omega t}\right)\nabla_H^2\zeta_0
 -\sigma\nabla_H^2\nabla_H^2\zeta_0 = 0.
\end{eqnarray*}
By substituting $w_0$ and $\zeta_0$ into the above equation, we find for
each harmonic $e^{ji\omega t/2}$,
\begin{equation}
  \begin{array}{rcl}
  H_1A_1 - fA_1^* - fA_3 &=& 0 \\
  H_3A_3 - fA_1 - fA_5 &=& 0 \\
  H_5A_5 - fA_3 - fA_7 &=& 0 \\
  &\vdots& \ \ ,
  \end{array}
  \label{eigen_zero}
\end{equation}
with
$
  H_j \equiv \left\{
      \rho\nu^2\left[ 4q_jk^4-k(q_j^2+k^2)^2 \right] - \rho gk^2 
        -\sigma k^4\right\}\big/{\textstyle{1\over 2}} \rho k^2.
$
By truncating the set of equations (\ref{eigen_zero}) at some $A_n$, 
the system can be solved numerically
as an eigenvalue problem.  This is indeed what was done by
Kumar and Tuckerman \cite{kumar94}.  However we observe that after
truncation at $A_n$, $A_n=fA_{n-2}/H_n,
A_{n-2}=fA_{n-4}/(H_{n-2}-{f^2\over H_n}),\cdots$. Therefore the 
set of equations can be rewritten as
$$
  \left( H_1 - { f^2 \over H_3 - {f^2 \over H_5- \cdots}}
  \right) A_1 - f A_1^* \equiv \bar H_1(k,f)A_1 - f A_1^* = 0,
$$
so that for a given wavenumber $k$, the threshold of instability 
$f_{0}$ is given by
$$
  f_0 = |\bar H_1(k,f_0)|.
$$
The complex amplitude $A_j$ can be recursively obtained from
Eq. (\ref{eigen_zero}) up to a real factor.  For an infinite system
the critical wavenumber $k_{\hbox{\scriptsize onset}}$ is the
wavenumber that corresponds to the lowest value of $f_0$.


Consider first the limit of low viscosity, and define
$\omega_0\equiv\omega/2$ as the time scale, and $k_0$ from $\omega_0^2
= gk_0 + \sigma k_0^3/\rho$ as the length scale, and the dimensionless
variables ${\bar k} = k/k_0, \gamma = 2\nu k_0^2/\omega_0,
G=gk_0/\omega_0^2, \Sigma = \sigma k_0^3/\rho\omega_0^2,$ and $\Delta
= f_0k_0/4\omega_0^2$.  For a damping coefficient $\gamma\ll 1$ and
$k$ near $k_{\hbox{\scriptsize onset}}$, $\Delta_{\hbox{\scriptsize
onset}}$ can be given explicitly as,
$$
  \Delta_{\hbox{\scriptsize onset}} 
= \gamma - {\textstyle{1\over2}}\gamma^{3/2} 
	+ {11-2G\over8(3-2G)}\gamma^{5/2} + \cdots, \label{threshold}
$$
with $0\leq G\leq 1$ by definition. The dimensionless critical
amplitude is proportional to $\gamma-{1\over2}\gamma^{3/2}$ at small
$\gamma$. While previous low damping calculations
\cite{milner91,zhang96} only used the linear term to determine the
location of the threshold, the first correction $-{1\over2}
\gamma^{3/2}$ can be a sizable contribution (e.g., a 15\% difference
at $\gamma=0.1$). We also note that a similar calculation for the
damped Mathieu equation leads to a threshold $\gamma + 3\gamma^2/64 +
O(\gamma^3)$, in which the first correction term is of a different
order and has a different sign.


To derive the amplitude equation we use the multiple scale approach
\cite{multiscale}. The solvability condition in this case arises from
the boundary conditions, not from the equation of motion as in most other
cases. The velocity field is expanded as,
$$
  {\bf u}=\epsilon^{1/2}{\bf u_0}+\epsilon{\bf u_1}+\epsilon^{3/2}{\bf u_2}
         +\cdots,
$$
with $\epsilon = (f-f_0)/f_0$, and similarly for $p$ and $\zeta$. Near
threshold, i.e., for $\epsilon\ll 1$, we separate fast and slow time
scales: $T=\epsilon t$;
$\partial_t\rightarrow\partial_t+\epsilon\partial_T$. Spatial slow
scales are not included because only regular patterns are considered
here. At order $\epsilon^{1/2}$ we recover the linear solution
discussed above.  Because we are interested in standing wave patterns
with different symmetries, the solution at this order is written as a
linear combination of waves with wavevectors ${\bf k}_m$ of magnitude
$k_{\hbox{\scriptsize onset}}$ in different directions on the $x$--$y$
plane,
$$
  w_0 = \sum_m\cos({\bf k}_m\cdot{\bf r})B_m(T)
        \sum_{j=1,3,5,\cdots} e^{ji\omega t/2} w_0^j(z) e_j 
                       +\hbox{c.c.}
$$
Here $B_m(T)$ are the {\em real} wave amplitudes, functions only of the
slow time scale $T$, and $e_j$ is $A_j$ found in Eq. (\ref{eigen_zero}).


At order $\epsilon$ the equation of motion for $w_1$ becomes
\begin{equation}
  \left(\partial_t\nabla^2 - \nu\nabla^2\nabla^2\right)w_1
     = [\nabla\times\nabla\times( {\bf u}_0 \cdot \nabla ) {\bf u}_0]_z.
  \label{firstorder}
\end{equation}
The solution $w_1$ contains terms of the form $\cos\left(({\bf
k}_m\pm{\bf k}_n)\cdot{\bf r}\right)$ that incorporate couplings
through stable modes, and that will contribute to the coefficients of
the cubic terms later in the expansion.  The particular solution
$w_{1\hbox{\scriptsize p}}$ is obtained by integrating
Eq. (\ref{firstorder}), with the homogeneous solution
$w_{1\hbox{\scriptsize h}}$ chosen so that the boundary conditions are
satisfied. Both the equation of motion and boundary
conditions at this order become very complicated. In order to find
$w_1$ in practice, we have developed a symbolic manipulation program
specific to this case, and found the solution on a computer.


At order $\epsilon^{3/2}$ the equation of motion becomes
\begin{eqnarray}
  & & \left(\partial_t\nabla^2 - \nu\nabla^2\nabla^2\right)w_2 
      \label{second_order_eq} \\
  &=&-\partial_T\nabla^2w_0 
   +\left\{\nabla\times\nabla\times[( {\bf u}_0 \cdot \nabla ) {\bf u}_1 
         +( {\bf u}_1 \cdot \nabla ) {\bf u}_0]\right\}_z . \nonumber
\end{eqnarray}
Only terms proportional to $\cos({\bf k}_1\cdot{\bf r})$ need
to be considered in the solution for $w_2$ and $\zeta_2$,
\begin{eqnarray*}
  w_2 &=& \cos({\bf k}_1\cdot{\bf r})
      \hskip -6pt\sum_{j=1,3,5,\cdots}\hskip -6pt
      e^{ji\omega t/2} 
      \left[E_j + \left(a_je^{kz} + b_je^{q_jz}\right)C_j\right] \\
  \zeta_2 &=& \cos({\bf k}_1\cdot{\bf r})
      \hskip -6pt\sum_{j=1,3,5,\cdots}\hskip -6pt
      e^{ji\omega t/2} C_j.
\end{eqnarray*}
Here $E_{j}(z)$ comes from the direct integration of
Eq. (\ref{second_order_eq}), and $a_je^{kz} + b_je^{q_jz}$ is the
homogeneous solution that has the same form as the linear solution.

Using the kinematic and tangential stress boundary conditions at order
$\epsilon^{3/2}$ we find (again with the symbolic manipulation
program) $a_j$ and $b_j$. The solution $w_2$ is finally inserted into
the normal stress boundary condition at order $\epsilon^{3/2}$ to
yield a system of equations for $C_j$ which
has the same left-hand side as
Eq. (\ref{eigen_zero}) but with nonzero right hand side.
Solving for $C_j$ just like in Eq. (\ref{eigen_zero}), we obtain
$$
  \bar H_1C_1 - f_0 C_1^* = F,
$$
with $F$ a function of the amplitude $B_m$. Since $f_0=|\bar H_1|$, by
requiring a nontrivial solution for $C_1$ we obtain the solvability
condition $F\bar H_1^*+F^*f_0=0$, which yields a standing wave
amplitude equation,
\begin{equation}
  {dB_1\over dT} = \alpha B_1 - g_0 B_1^3
   - \sum_{m\not=1}g(\theta_{m1})B_m^2B_1,
  \label{amplitude_equation}
\end{equation}
with $\theta_{m1}$ the angle between ${\bf k}_m$ and ${\bf k}_1$, and explicit
expressions for the coefficients.
%
%
Equation (\ref{amplitude_equation}) is of gradient form, and can be derived
from a Lyapunov function
\begin{equation}
\label{eq:lyapunov}
  {\cal F} =-{1\over2}\alpha\sum_mB_m^2
            +{1\over4}\sum_m\sum_n g(\theta_{mn})B_m^2B_n^2,
\end{equation}
which may be used to find the preferred pattern near threshold
\cite{review}. For regular patterns of $N$ standing waves, ${\bf k}_m$
form a regular polygon, and the $B_m$ are constant.


We next turn to a comparison between the selected patterns predicted
by Eq. (\ref{eq:lyapunov}), and two recent sets of systematic
experimental surveys involving large aspect ratio systems, both of
which aiming at addressing the issue of pattern selection over a wide
range of experimental parameters \cite{kudrolli96a,binks96}.  Binks
and van de Water \cite{binks96} have focused on a low viscosity fluid
\cite{fluid_prop}, a large aspect ratio cell, and a layer depth much
larger than the wavelength.  When the driving frequency is decreased
from 45Hz, a transition from a $N=2$ square pattern to a $N=3$
hexagonal pattern was observed at approximately 35Hz, and to a
quasi-periodic $N=4$ eight-fold pattern at approximately 29Hz
\cite{observed_pen}. Our prediction for these transitions based on
Eq. (\ref{eq:lyapunov}) are 35.4~Hz and 28.7~Hz respectively. These
results are also in good agreement with the earlier weak damping
calculation \cite{zhang96}
that predicted the same transitions at frequencies of 32.8~Hz and
27.9~Hz respectively.

\begin{figure}
\psfig{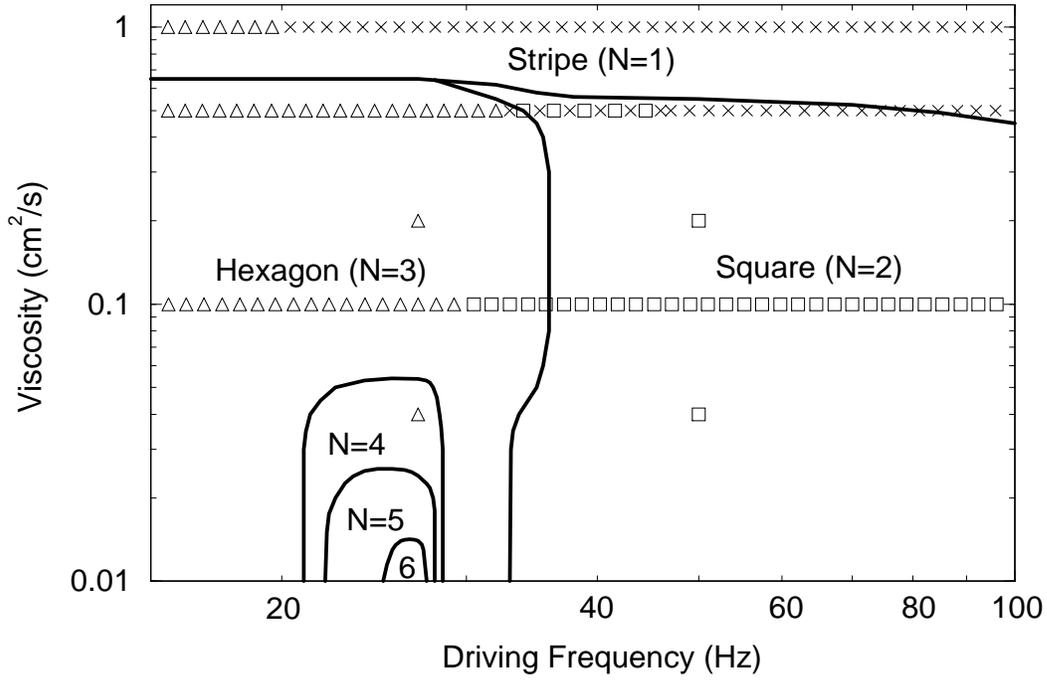}
  \caption{Preferred patterns in viscosity-driving frequency space.
           Symbols represent the experimental results.
           $\times$=stripe, $\Box$=square, and $\triangle$=hexagon.
	   Alternating $\times$ and $\Box$ indicate mixed-stripe-square 
	   states.}
   \label{fi:selected}	   
\end{figure}

A large aspect ratio experiment involving fluids of various
viscosities has been carried out by Kudrolli and Gollub
\cite{kudrolli96a}. Although the fluid depth (0.3~cm) is smaller than
the wavelength in the experiment (1--3~cm), the comparison is still
illuminating. Figure \ref{fi:selected} shows the symmetry of the
preferred patterns predicted by our calculations in the parameter
space defined by the viscosity of the fluid and the driving frequency
(with $\rho=0.95\hbox{g}/\hbox{cm}^3$ and
$\sigma=20.6\hbox{dyne}/\hbox{cm}$), and the experimentally observed
patterns. Stripe patterns are preferred at high viscosity, whereas at
low viscosity, hexagons (at low frequency) and squares (at high
frequency) are observed. We also show the small region in which a
sequence of quasiperiodic patterns are expected to be selected. The
experimental results by Kudrolli and Gollub are shown as the symbols
in the figure. We note an excellent agreement in the regions in which
stripes, squares, and hexagons are observed, specially in view of the
shallow fluid depth involved in the experiment. The shallowness of the
layer probably
accounts for the observation of a hexagonal pattern at
$\nu=1\hbox{cm}^2/\hbox{s}$ and low frequency, and not observing a
quasiperiodic pattern for $\nu=0.04\hbox{cm}^2/\hbox{s}$ and
$f=27\hbox{Hz}$. As noted above, the experiments by Binks and van de 
Water \cite{binks96} did probe this latter region in a deep 
fluid layer, with their results agreeing with our predictions.


In summary, we have presented a nonlinear theory for Faraday waves in 
viscous fluids with no assumptions or approximations other than those
inherent to the multiscale expansion. A set of standing wave amplitude
equations has been obtained that is of gradient form. Minimization of the
associated Lyapunov function leads to determination of the preferred
pattern near threshold. The predicted patterns are in excellent 
agreement with recent experiments in large aspect ratio systems involving
a range of fluid viscosities and driving frequencies.


This research has been supported by the
U.S. Department of Energy, contract No. DE-FG05-95ER14566, and also
in part by the Supercomputer Computations Research Institute, which is
partially funded by the U.S. Department of Energy, contract No.
DE-FC05-85ER25000.


\begin{references}
\bibitem{fluid_prop} The fluid properties are:
  $\nu=0.03397\hbox{cm}^2/\hbox{s}$ ($\gamma\sim 0.01 - 0.03$),
  $\rho=0.8924\hbox{g}/\hbox{cm}^3$, and $\sigma=18.3\hbox{dyne}/\hbox{cm}$.
\bibitem{observed_pen} In the experiment, patterns with $N=5$ are observed
   around 27Hz. As it is also argued by the authors of the experiment,
   this could be attributed to finite size effects since ${\cal F}$ 
   for $N = 5$ is larger but very
   close to ${\cal F}$ for $N=4$. The difference at about 26~Hz is less than
   0.2\%.
\end{references}
\end{document}